\documentclass[fleqn,12pt,twoside]{article}
\usepackage{espcrc1}

\usepackage{graphicx}
\usepackage[figuresright]{rotating}

\newcommand{\AmS}{{\protect\the\textfont2
  A\kern-.1667em\lower.5ex\hbox{M}\kern-.125emS}}
\def\5LHe{$^5_\Lambda$He}

\def\12LC{$^{12}_\Lambda$C}

\def\piK{$(\pi^+, K^+)$}
\def\Kpi{$({K}^-, \pi^-)$}
\def\Lppi-{$\Lambda \to p\pi^-$}
\def\Lnpi0{$\Lambda \to n\pi^0$}
\def\Lpnp{$\Lambda p\to np$}
\def\Lnnn{$\Lambda n\to nn$}

\def\6LipiK{$^6$Li($\pi^+, K^+$)}
\def\6LipiK6LLi{$^6\mathrm{Li}(\pi^+, K^+)^6_\Lambda$Li}
\def\6LipiKp5LHe{$^6\mathrm{Li}(\pi^+, K^+p)^5_\Lambda$He}

\title{Measurement of the $\pi^-$ decay width of $^5_\Lambda$H\lowercase{e}}

\author
{ S.~Kameoka$^a$,  
  S.~Ajimura$^b$,
  K.~Aoki$^c$,
  A.~Banu$^d$,
  H.~C.~Bhang$^e$,
  T.~Fukuda$^f$,
  O.~Hashimoto$^a$,  
  J.~I.~Hwang$^e$,
  B.~H.~Kang$^e$,
  E.~H.~Kim$^e$,
  J.~H.~Kim$^e$,
  M.~J.~Kim$^e$,
  T.~Maruta$^h$,  
  Y.~Miura$^a$,
  Y.~Miyake$^b$,
  T.~Nagae$^c$,
  M.~Nakamura$^h$,  
  S.~N.~Nakamura$^a$,
  H.~Noumi$^c$,
  S.~Okada$^g$,
  Y.~Okayasu$^a$,
  H.~Outa$^i$,
  H.~Park$^j$,
  P.~K.~Saha$^f$,
  Y.~Sato$^c$,
  M.~Sekimoto$^c$,
  T.~Takahashi$^a$,
  H.~Tamura$^a$,
  K.~Tanida$^i$,
  A.~Toyoda$^c$,
  K.~Tsukada$^a$,
  T.~Watanabe$^a$,
  H.~J.~Yim$^e$ \\
  \vspace{2mm}
  $^a$Department of Physics, Tohoku University, Miyagi, 980-8578, Japan \\
  $^b$Department of Physics, Osaka University, Osaka, 560-0043, Japan \\
  $^c$High Energy Accelerator Research Organization (KEK),
  Ibaraki, 305-0801, Japan \\
  $^d$Gesellschaft f$\ddot{\mbox{u}}$r Schwerionenforschung mbH (GSI),
  Darmstadt, 64291, Germany \\
  $^e$Department of Physics, Seoul National University, Kwanak-gu,
  151-742, Korea \\  
  $^f$Osaka Electro-Communication University, Osaka, 572-8530, Japan \\
  $^g$Department of Physics, Tokyo Institute of Technology,
  Tokyo, 152-8551, Japan \\
  $^h$Department of Physics, University of Tokyo, Tokyo, 113-0033, Japan \\
  $^i$The Institute of Physical and Chemical Research (RIKEN),
  Saitama, 351-0198, Japan \\
  $^j$Korea Research Institute of Standards and Science (KRISS),
  Daejeon, 305-600, Korea
  }
\begin{document}

\maketitle

\begin{abstract}
 We have precisely measured $\Lambda \to p\pi^-$ decay width of \5LHe
 and demonstrated significantly larger
 $\alpha$ -$\Lambda$ overlap than expected from
 the central repulsion $\alpha$-$\Lambda$ potential,
 which is derived from YNG $\Lambda$-nucleon interaction.
\end{abstract}

 \section{Introduction}
 
 It is known that hypernuclei mainly decay via mesonic decay(\Lppi-,
 \Lnpi0) or non-mesonic decay(\Lnnn, \Lpnp).
 Mesonic decay releases only some 40~MeV energy
 and undergoes Pauli suppression by the existence of other nucleons. 
 Thus the mesonic decay rate is sensitive to the overlap of $\Lambda$
 wave function with nucleus, and reflects nucleus-$\Lambda$ potential
 shape.

 Motoba {\it et al.}\ constructed a $\Lambda$ single particle potential
 in \5LHe employing realistic $\Lambda N$ interaction ``YNG'' with folding
 procedure.\cite{mot}
 This potential is characterized by a repulsive core, 
 which has the effect to push $\Lambda$ wave function toward the
 outside of nucleus and leads to an enhancement of the mesonic decay
 rate by the relaxation of Pauli suppression.
 They calculated mesonic decay rate of \5LHe with the $\Lambda$ wave
 function as an initial state, 
 and compared it to the one derived from one-range Gaussian $\Lambda N$
 interaction ``ORG''.
 The \Lppi- decay width showed about 30\% difference: 
 $\Gamma_{\pi^-}(\mathrm{ORG})=0.321~\Gamma_\Lambda$,
 $\Gamma_{\pi^-}(\mathrm{YNG})=0.393~\Gamma_\Lambda$,
 where $\Gamma_\Lambda$ denotes total decay width of free $\Lambda$.
 Kumagai-Fuse {\it et al.}\ also performed the same kind of calculation and
 reported similar results for \5LHe.\cite{kum}
 However, existing experimental data,
 $\Gamma_{\pi^-}(\mathrm{Exp.})=0.44\pm0.11~\Gamma_\Lambda$,
 has large error and cannot distinguish above two potentials.\cite{szy}
 It should be noted that both of the interactions are determined to
 reproduce $\Lambda$ binding energy of $^5_\Lambda$He, and could not be
 distinguished without measuring the difference in the mesonic decay rate.
 Therefore, precise measurement of pionic decay rate is
 important to check the validity of the YNG interaction.

 The $\pi^-$ decay width is obtained by dividing the
 branching ratio by the lifetime.
 Thus, both of these observables need to be measured with high precision.

 \section{Experiment and analysis}
 
 The experiment was performed using K6 beam line of KEK 12~GeV
 proton synchrotron.
 $^6$Li target was bombarded with 1.05~GeV/c $\pi^+$ beam and
 \5LHe hypernucleus were produced via
 $^6\mathrm{Li}(\pi^+, K^+)^6_\Lambda$Li,
 $^6_\Lambda\mathrm{Li}\to ^5_\Lambda\mathrm{He}+p$ reaction.
  The incoming pion and outgoing kaon momenta were measured by means of
  magnetic spectrometers and missing mass was reconstructed.
  A couple of detector system was installed around the target
  so as to measure trajectories and velocities of hypernuclear decay
  products.
  The decay counter comprises 4~mm-thick plastic scintillators named T2,
  2~cm-thick ones named T3,
  six layers of 5~cm-thick ones named T4,
  and a drift chamber placed between T2 and T3.

  \subsection*{$\pi^-$ decay branching ratio}
  
  Figure~\ref{pid} shows PID function of charged particles emitted from
  the decay of \5LHe,
  which is derived from particle velocity, $dE/dx$ and total energy
  deposit in sequentially fired scintillators.
  A tail on positive side of the pion peak is attributed to an
  extra energy deposit caused by nuclear pion absorption.
  
  Figure~\ref{mhyp} shows the excitation spectra of $^6$Li$(\pi^+, K^+)$
  reaction.
  Hypernuclear formation events were successfully separated as a peak
  structure in both (up) inclusive and (down) pion coincidence spectra.
  The region between dotted lines represents the events accepted as
  \5LHe production.
  Use of \piK reaction enabled
  low background hypernuclear spectroscopy,
  while the previous measurement using \Kpi reaction
  suffers serious background from beam kaon decay and could not avoid large
  ambiguity in determination of $\pi^-$ emission ratio.

  We cannot well identify
  such decay particles as have large angles against vertical axis
  and pass through side of T4 array without losing whole kinetic energy.
  Our analysis required two(three) segments at the center of T2(T3) counter
  to be fired so as to avoid such ambiguity and
  accurately determine pion detection efficiency.

  Electrons created by \Lnpi0 decay followed by electro-magnetic shower
  cannot be distinguished from pion.
  The electron contamination rate was evaluated by Monte Carlo method
  using GEANT and effectively subtracted.\cite{gea}

  The pion emission rate is stable when the hypernuclear mass gate is
  narrowed, 
  which confirms that contamination with higher excited states is
  negligible.

  $\pi^-$ decay branching ratio was
  determined to be $b_{\pi^-}=0.359\pm 0.009$ (statistical error only).

%
\begin{figure}[ht]
 \begin{minipage}{0.44\linewidth}
  \begin{center}
   \includegraphics[width=\linewidth]{pid.epsi}
   \caption{Charged particle identification emitted from the decay of \5LHe. \label{pid}}
  \end{center}
 \end{minipage}
\begin{minipage}{0.02\linewidth}
 ~
\end{minipage}
\begin{minipage}{0.54\linewidth}
  \begin{center}
   \includegraphics[width=\linewidth]{mhyp.epsi}
   \caption{the excitation spectra of $^6$Li($\pi^+, K^+$)
   reaction for (up) inclusive and (down) pion coincident
   events. \label{mhyp}}
  \end{center}
\end{minipage}
 \end{figure}
%
%

\subsection*{Lifetime}

The hypernuclear decay time was directly measured as
$\Delta t = t_{T2} - l_{d}/v_{d}  - l_{b}/v_{b} - t_{T1}$,  
where $t_{T2}$ was the time when the decay particles reached T2 counter, 
$l_{d}$ is flight length of the decay particles from target to T2
counter, $v_{d}$ is velocity of the decay particles, $l_{b}$ is flight
length of beam pion from T1 counter to target, $v_{b}$ is velocity of
beam pion, and $t_{T1}$ is beam injection time at T1 counter.
The lifetime was derived by fitting the decay time spectrum by
exponential function convoluted with time response function of the
detector system, which is obtained as $\Delta t$ spectrum for
prompt reaction caused via strong interaction.\cite{par}
$(\pi, pp)$ events, which means that a scattered proton was detected by
SKS and another proton was detected by the decay counter,  
were simultaneously collected for this purpose and 
fluctuation of the time response function was corrected every two hours
during data acquisition. 
Figure~\ref{life}(right) shows prompt(up) and delayed(down) time spectra.
With above procedure, lifetime of \5LHe was determined to be $\tau =
278^{+11}_{-10}$~ps.

Lifetime of $^{12}_\Lambda$C was also measured as $212\pm 6$~ps,
whose value is twice more accurate over the previous measurement.\cite{par}
See Figure~\ref{life}(left) for the fitting of $^{12}_\Lambda$C lifetime.
\begin{figure}[ht]
 \begin{minipage}{0.49\linewidth}
  \begin{center}
   \includegraphics[width=\linewidth]{life.epsi}
  \end{center}
 \end{minipage}
 \begin{minipage}{0.02\linewidth}
  ~
 \end{minipage}
 \begin{minipage}{0.49\linewidth}
  \begin{center}
   \includegraphics[width=\linewidth]{life_E508.epsi}
  \end{center}
 \end{minipage}
 \caption{
 Left: Prompt time spectrum of  $^6\mathrm{Li}(\pi^+, pp)$ events (up)
 and decay time spectrum of \5LHe (down).
 Right: Prompt time spectrum of $^{12}\mathrm{C}(\pi^+, pp)$ events (up)
 and decay time spectrum of \12LC (down).\label{life}
 }
\end{figure}

\section{Results and discussion}

 The total decay width and $\pi^-$ decay width of \5LHe obtained were 
 $\Gamma_\mathrm{tot}=1/\tau =0.947\pm 0.038~\Gamma_{\Lambda}$ and
 $\Gamma_{\pi^-}=b_{\pi^-}\times \Gamma_\mathrm{tot}=0.340\pm
 0.016~\Gamma_{\Lambda}$,  respectively.

 Table~\ref{tab1} compares the present experiment with previous
 one\cite{szy} and the theoretical calculation based on ORG/YNG
 $\alpha$-$\Lambda$ potential.
 One can see that data accuracies were drastically improved although
 further careful check for systematic error is remained to be done. 
 Our measurement shows that $\pi^-$ decay width is considerably smaller
 than the theoretical prediction using YNG $\Lambda N$ interaction.
 This suggests that new $\alpha$-$\Lambda$ potential such as gives
 larger overlap between $\Lambda$ and nucleus need to be considered.
 \newpage
%
 %
 \begin{table}[h]
  \begin{center}
   \begin{tabular}{c|c|c|c|c}
    \hline
    {} & \multicolumn{2}{c|}{Experiment} &
    \multicolumn{2}{c}{Theory\cite{mot}}\\
    \cline{2-5}
    {} & present & previous\cite{szy} & ORG & YNG\\
    \hline\hline
    {} &{} &{} &{} &{}\\[-2ex]
    $\Gamma_\mathrm{tot}/\Gamma_\Lambda$ &$0.947\pm$ 0.038 &$1.03\pm 0.08$ &--- &---\\[0.5ex]
    \hline
    {} &{} &{} &{} &{}\\[-2ex]
    $\Gamma_{\pi^-}/\Gamma_\Lambda$ &$0.340\pm 0.016$ &$0.44\pm 0.11$ &0.321 &0.393\\[0.5ex]
    \hline
   \end{tabular}\label{tab1}
   \caption{The experimental results compared with previous experiment
   and theoretical calculations. The errors of the present data are
   statistical only.}
  \end{center}
 \end{table}


\begin{thebibliography}{0}
 \bibitem{mot} T. Motoba {\it et al.}, {\it Nucl. Phys.}  
	 {\bf A534}, 597 (1994).

 \bibitem{kum} I. Kumagai-Fuse {\it et al.}, {\it Phys. Lett.}
	 {\bf B345}, 386 (1995).

 \bibitem{szy} J. J. Szymanski {\it et al.}, {\it Phys. Rev.}
	 {\bf C43}, 849 (1991).
	 
 \bibitem{par} H. Park {\it et al.}, {\it Phys. Rev.}
	 {\bf C61}, 054004 (2000).

 \bibitem{gea} CERN Program Library, GEANT.
	 
\end{thebibliography}
\end{document}